\documentclass[10pt,conference]{IEEEtran}
\IEEEoverridecommandlockouts
\usepackage{cite}
\usepackage{amsmath,amssymb,amsfonts}
\usepackage{algorithmic}
\usepackage{graphicx}
\usepackage{textcomp}
\usepackage{xcolor}

\usepackage{xspace}
\usepackage{url}
\usepackage[hidelinks]{hyperref}
\usepackage[inline,shortlabels]{enumitem}
\usepackage{listings}

\definecolor{darkgreen}{rgb}{0,0.6,0}
\newcommand{\cf}{\textit{cf.}}
\newcommand{\exciting}{{\tt\textbf{exciting}}\xspace}

\newcommand{\Nomad}{NOMAD\xspace}
\newcommand{\code}{{\tt code}\xspace}
\newcommand{\Code}{{\tt Code}\xspace}
\newcommand{\codes}{{\tt codes}\xspace}
\newcommand{\Codes}{{\tt Codes}\xspace}
\newcommand{\codeBlue}[1]{{\textcolor{blue}{#1}}}
\newcommand{\codeGreen}[1]{{\textcolor{darkgreen}{#1}}}

\lstdefinestyle{XML}{
	basicstyle=\scriptsize,
	numbers=left,
	numbersep=5pt,
	numberstyle=\tiny\color{black},
	stepnumber=1,
	keywords=[1]{input, title, structure, crystal, basevect, species, atom, dfthalfparam, shell, groundstate, dfthalf},
	keywordstyle=[1]\color{blue},
	keywords=[2]{speciespath, scale, rmt, coord, cut, ampl, exponent, number, ionization, do, rgkmax, gmaxvr, ngridk, outputlevel, xctype, printVSfile},
	keywordstyle=[2]\color{darkgreen},
	xleftmargin=15pt,
	xrightmargin=15pt,
}

\usepackage{url}
\makeatletter
\def\ps@IEEEtitlepagestyle{
	\def\@oddfoot{\mycopyrightnotice \thepage}
	\def\@evenfoot{}
}
\def\mycopyrightnotice{
	{\footnotesize
		\begin{minipage}{\textwidth}
			\centering
			\textcopyright~2019 IEEE. Personal use of this material is permitted.
			Permission from IEEE must be obtained for all other uses, in any current or future media, including reprinting/republishing this material for advertising or promotional purposes, creating new collective works, for resale or redistribution to servers or lists, or reuse of any copyrighted component of this work in other works. 
			{\tt DOI:} \url{https://doi.org/10.1109/SE4Science.2019.00010}
		\end{minipage}
	}
}
\hypersetup{
	bookmarks=true,         
	unicode=false,          
	pdftoolbar=true,        
	pdfmenubar=true,        
	pdffitwindow=false,     
	pdftitle={Challenges for Verifying and Validating Scientific Software in Computational Materials Science},
	pdfauthor={Thomas Vogel, Stephan Druskat, Markus Scheidgen, Claudia Draxl, Lars Grunske},
	pdfsubject={},
	pdfcreator={},
	pdfproducer={},
	pdfkeywords={Verification and Validation, Scientific Software, Computational Materials Science},
	pdfnewwindow=true,
}
\def\BibTeX{{\rm B\kern-.05em{\sc i\kern-.025em b}\kern-.08em
    T\kern-.1667em\lower.7ex\hbox{E}\kern-.125emX}}
\begin{document}

\title{Challenges for Verifying and Validating Scientific Software in Computational Materials Science}

\author{
	\IEEEauthorblockN{
		Thomas Vogel\IEEEauthorrefmark{1},
		Stephan Druskat\IEEEauthorrefmark{1}\IEEEauthorrefmark{3},
		Markus Scheidgen\IEEEauthorrefmark{2},
		Claudia Draxl\IEEEauthorrefmark{2},
		and
		Lars Grunske\IEEEauthorrefmark{1}
	}
	\IEEEauthorblockA{\IEEEauthorrefmark{1}\textit{Computer Science Department}, \textit{Humboldt-Universit\"at zu Berlin}, Berlin, Germany}
	\IEEEauthorblockA{\IEEEauthorrefmark{3}\textit{German Aerospace Center (DLR)}, Berlin, Germany\\
		Email: thomas.vogel@informatik.hu-berlin.de, stephan.druskat@dlr.de, grunske@informatik.hu-berlin.de}
	\IEEEauthorblockA{\IEEEauthorrefmark{2}\textit{Physics Department and IRIS Adlershof}, \textit{Humboldt-Universit\"at zu Berlin}, Berlin, Germany\\
		Email: markus.scheidgen@physik.hu-berlin.de, claudia.draxl@physik.hu-berlin.de}
}

\maketitle
\pagestyle{plain}
\begin{abstract}
Many fields of science rely on software systems to answer different research questions.
For valid results researchers need to trust the results scientific software produces, and
consequently quality assurance is of utmost importance. In this paper we are investigating
the impact of quality assurance in the domain of computational materials science (CMS). Based on our experience
in this domain we formulate challenges for validation and verification of scientific software
and their results. Furthermore, we describe directions for future research that can potentially
help dealing with these challenges.
\end{abstract}


\section{Introduction}

\noindent
Software has become an important driver for research in many scientific disciplines such as biology and physics~\cite{Carver2016se4sience}. 
Scientists often use software in experiments to produce evidence for the validity of their theories, and publish scientific papers based on this evidence~\cite{SandersK08}.
However, in the worst case the validity of such a computational experiment -- and thus of the (published) research results -- may be jeopardized if the software producing the evidence is not of sufficient quality. A software that has bugs may produce wrong data leading to erroneous evidence.
Accordingly, scientific papers have been retracted in the past due to issues with software~\cite{Miller1856}.

Consequently, software engineering principles are being increasingly adopted~\cite{Hannay2009,Nguyen-HoanFS10,HeatonC15,Storer17,JohansonH18},
and best practices for scientific software development processes have been proposed~\cite{KellyHS09,Wilson2014}.
At the same time, a clash of cultures between software engineers and domain scientists has been reported~\cite{Segal05,Segal08}.

In this context, validation and verification of scientific software are critical, as they establish trust in the software for it to perform the required calculations correctly. %
In this regard, inadequate behavior of scientific software is a threat to the validity of research results, and has consequently been a main subject of research~\cite{HattonR94}.
To demonstrate the correctness of scientific software, testing is considered essential~\cite{Clune2014testing}, and has been investigated for scientific software~\cite{SandersK08,KellyGS11,KellyTH11,LaneG12,Kelly15,Kanewala2014}, resulting in tools for testing scientific software~\cite{SmithKRY04,Dubois12},
the beneficial use of reference data for testing~\cite{CoxH1999}, and
test-driven development methods~\cite{Nanthaamornphong2017,CluneR11}.
Despite these advances in testing scientific software, all approaches suffer from the \textit{oracle problem} and \textit{large variability} (i.e., a large configuration space and input domain) of the software under test~\cite{Kanewala2014}.
Carver et al.~\cite[p.\,554]{CarverKSP07} faced the oracle problem in five case studies of computational science and engineering projects, and concluded: ``Validation is problematic because it is often difficult, or even impossible, to establish the correct output or result a priori.''
In contrast, testing from a software engineering perspective typically considers accurate oracles, that is, the expected output of the software under test is precisely known. This results in a binary oracle: The calculated output either does or does not match the expected output. This contradicts the nature of scientific software, where oracles are unknown or not precisely known.
Moreover, the large variability of scientific software poses a challenge to standard testing tools from software engineering because of the large number of tests that are required to comprehensively test the software. Consequently, tests should be well chosen with the goal of allowing scientists to increase their trust in the software~\cite{HookK09}.

In this paper, we investigate the validation and verification of scientific software in computational materials science (CMS). CMS is concerned with the design and discovery of new materials using computational methods.
Based on our experience in the CMS domain, we discuss corresponding challenges such as (i)~the oracle problem, and (ii)~large configuration spaces of CMS programs, called \codes, taking the specifics of the domain into account. In the context of the development and use of the \Nomad~\cite{DraxlS2018} ecosystem of \codes and data, we further discuss challenges related to (iii)~large-scale, heterogeneous data, and (iv)~global software development.
Corresponding to these challenges, we proceed to discuss directions for future research on validating and verifying scientific software in CMS.

Throughout the paper, we take the perspective of a CMS scientist who runs calculations to design and analyze materials using a \code such as \exciting~\cite{exciting}, ABINIT~\cite{abinit}, or VASP\footnote{\url{https://www.vasp.at/}}, or a data-analysis workflow in \Nomad. As the results of a calculation rely on the validity and correctness of the used \code, our goal is to derive trust levels for \codes from testing, so that the scientist can increase her trust in the \code. This paves the way for trustworthy, reproducible calculations and research results.

\section{Computational Materials Science}
\label{sec:cms}

\noindent
The convergence of theoretical physics and chemistry, materials science and engineering, and computer science into \emph{computational materials science} (CMS) enables the modeling of materials (both existing materials and those that can be created in the future) at the electronic and atomic level. This allows the accurate prediction of how these materials will behave at the microscopic and macroscopic levels, and of their suitability for specific research and commercial applications.
CMS is characterized by a healthy, but heterogeneous ecosystem of many different CMS programs, called \codes, developed by different research groups across the globe. These \codes are highly domain-specific scientific software packages implementing various theoretical methods. They are executed in high performance computing centers, with millions of CPU hours spent every day, some of them at petascale performance, producing a large stock of equally heterogeneous CMS data.

The \Nomad Center of Excellence\footnote{\url{https://nomad-coe.eu/}} (EU/Horizon 2020) aims to enable the CMS community to provide CMS data along the FAIR (findable, accessible, interoperable, and re-usable \cite{wilkinson_fair_2016}) principles of data sharing. The \Nomad platform provides services that allow scientists to upload raw \code inputs and outputs and to automatically convert data from all relevant \codes into a \code-independent normalized format. It further allows scientists at various levels of expertise to search, inspect, analyze, and visualize all data in this \code-independent format. Currently, \Nomad supports over 40~\codes, and stores more than 50 million results of complex calculations regarding properties of materials, including those of the largest US databases,
provided by several hundred individual researchers and research groups. Its \code-independent format uses a hierarchical data schema with over 400 common \code-independent and almost 2.000 \code-specific attributes.

\begin{figure}[t]
	\centering
	\includegraphics[width=1\columnwidth]{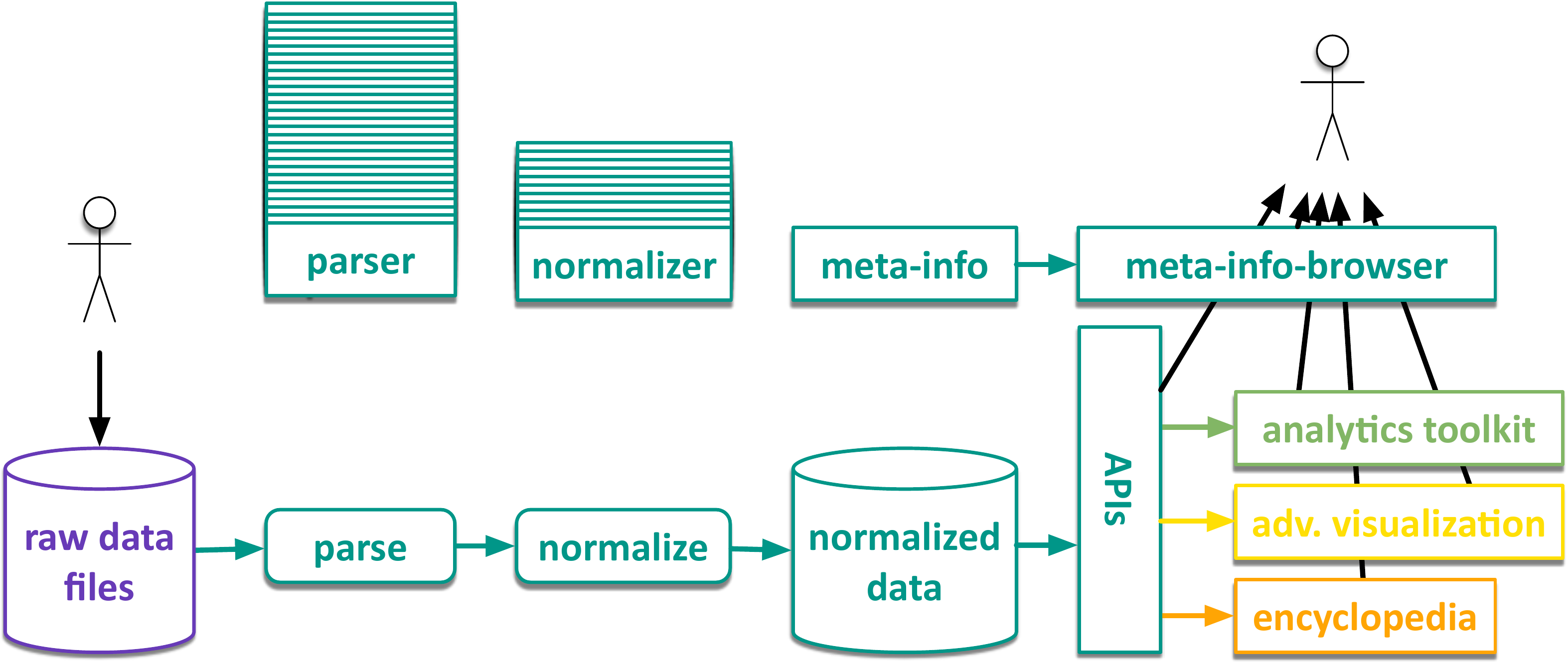}
	\caption{Architectural view of \Nomad.}
	\label{figure:nomad_architecture}
	\vspace{-1.5em}
\end{figure}

The architecture of the \Nomad platform (see Fig.~\ref{figure:nomad_architecture}), consists of six major components:
\begin{enumerate*}
	\item The \textit{raw data files} Repository where scientists upload, search, and download raw data.
	\item \textit{Parsers} and \textit{normalizers} that convert raw data in a \code-specific format to so-called \emph{Archive} data whose format is \code-independent.
	\item The Archive data, that is, the \textit{normalized data} that can be accessed through an API.
    \item The \emph{Analytics Toolkit} that allows scientists to apply machine learning techniques to CMS data.
    \item The \emph{Encyclopaedia} that aggregates calculations to provide a comprehensive and consistent collection of data for all materials.
    \item The \emph{advanced visualization} that uses 3D and virtual-reality techniques to visualize materials at an atomic level.
\end{enumerate*}

Following an hour-glass model, the most crucial part of \Nomad is the Archive of normalized data. All supported 40+~\codes use a different format to represent input and output data for their individual simulations/calculations. \Codes, data, and data formats differ in the following aspects.
First, \codes implement different methods, with varying computational parameters -- and thus, numerical precision -- and individual limitations and trade-offs.
Second, \codes focus on different aspects and produce different physical properties of a simulated material. For instance, a \code may specialize in electrical, optical, or thermal properties.
Third, data is provided in different unit systems (e.g., International System Units (SI) or atomic units).
Fourth, although most \codes use a text format that adheres to some community standards, all quantities are presented in different orders, and matrices and vectors are laid out differently. Quantity values range from strings and dates, simple numerical values, to large vectors, matrices, and tensors of several GB, or even TB, in size. Data formats are not formalized, and documentation is often sparse. Data of individual calculations is often spread over multiple files. Relations between calculations may exist as typically one calculation is based on another. However, such relations are not formalized and have to be deduced from common practices, for instance, a commonly used layout of directories.

To represent data in a normalized and homogeneous form, \Nomad defines an ontology-like data model that unifies all \codes with a common schema. The schema is used to formalize, categorize, and document all \codes, as well as \code-common and \code-specific quantities, in a single evolving model, called \emph{meta-info}. It uses a proprietary schema language
that specializes in describing physical quantities (e.g., with units and vector/matrix dimensions).
\emph{meta-info} is independent of distinct technical data formats, and the Archive data can be represented in different technical file formats. For example, \Nomad stores the archive data in HDF5, but the API supports access to the data via a JSON representation.

To convert raw CMS data to Archive data, \Nomad uses 40+ parsers (one per \code) and several normalizers. Each set of \code input/output data is parsed and then processed by all normalizers. %
Parsers re-produce all quantities found in the raw data in their respective \emph{meta-info} form. Normalizers then compute derived properties, classify simulations, convert units, and relate data with other sources (e.g., external materials databases). In computer language terms, parsers and normalizers only work on a syntactical level, all semantics is added by other \Nomad and potential third-party services.

One of the 40+ \codes used in the context of \Nomad is \exciting, a software package implementing density-functional theory (DFT) and many-body perturbation theory~\cite{exciting}.
As suggested by its name, \exciting has a major focus on calculating excited-state properties of materials.

\section{Problem Statement and Challenges}
\label{sec:problem+challenges}

\noindent
In this section, we first discuss the problem statement, including its relevance to scientific software with a focus on \codes in computational materials science (CMS). We then proceed to detail challenges in verifying and validating such software.

\subsection{Problem Statement}
\label{subsec:problem}

\noindent
To design and discover new materials, CMS scientists conduct \textit{computational experiments}, in which large-scale, heterogeneous data is processed by data-analysis workflows including \codes.
While certain steps of a workflow are concerned with preparing data (e.g., parsers), the \codes perform scientific calculations.
For an experiment, scientists reuse existing \codes and data, as well as develop new \codes that produce new data. Subsequently, they combine all elements in a workflow for execution.
The results of such experiments often provide the evidence that the scientists' theories work, which is consequently the basis for scientific papers.
A recent example from CMS is the work by Rodrigues Pela et al.~\cite{RodriguesPela+2018} who use \exciting to perform all required calculations.

In this context, there is a multitude of \codes in CMS implementing a range of theoretical methods. Each of these methods relies on a set of computational parameters that govern the numerical precision of the respective implementation of the method. Choosing the best \code, and optimal parameters that guarantee high precision is a non-trivial issue, which greatly influences the calculation results. Similar observations of configuration choices impacting research results are reported in bioinformatics~\cite{Cashman:2018}. This large configuration space results in high variability, which challenges CMS scientists in predicting how configuration choices impact the results.

Consider, for example, the basic input file for \exciting in Listing~\ref{listing:input}. This input file is used for DFT-1/2 calculations (specifically, LDA-1/2) for silicon, in order to compute single-particle band gaps. It first defines the \codeBlue{title} (line~2) and the material \codeBlue{structure}, in this case silicon (lines 3--11). It also features several parameters, such as the presence of \codeBlue{dfthalf} (see line~19), which triggers the DFT-1/2 calculation rather than the default standard DFT. The DFT-1/2 method is configured by the parameters in lines~12--14. Certain parameters have to be determined variationally to obtain optimal results (e.g., \codeGreen{cut}) while others are constant (e.g., \codeGreen{exponent} is usually set to 8).\footnote{For further details, see \url{http://exciting-code.org/nitrogen-dft05}.}

\begin{lstlisting}[caption=Example of an input for \exciting., label=listing:input, style=XML, belowskip=1\baselineskip]
<input>
   <title>Bulk Silicon: LDA-1/2 example</title>
   <structure speciespath="$EXCITINGROOT/species">
      <crystal scale="10.26">
         <basevect>0.0   0.5   0.5</basevect>
         <basevect>0.5   0.0   0.5</basevect>
         <basevect>0.5   0.5   0.0</basevect>
      </crystal>
      <species speciesfile="Si.xml" rmt="2.1">
         <atom coord="0.00  0.00  0.00"></atom>
         <atom coord="0.25  0.25  0.25"></atom>
         <dfthalfparam cut="3.90" ampl="1" exponent="8">
            <shell number="0" ionization="0.25" />
         </dfthalfparam>
      </species>
   </structure>
   <groundstate do="fromscratch" rgkmax="7.0" gmaxvr="14"
      ngridk="6 6 6" outputlevel="high" xctype="LDA_PW">
      <dfthalf printVSfile="false"/>
   </groundstate>
</input>
\end{lstlisting}

This (extremely simple) example illustrates only a very small fraction of the complexity of configuring \codes, as there exist configuration parameters that are mutable or constant, as well as dependencies between parameters (e.g., only if the \codeBlue{dfthalf} method is selected, it can be further configured by the parameters in \codeBlue{dfthalfparam}).

Consequently, given the complexity of configuring \codes for their use, and of initially developing such \codes, scientists may generally ask themselves two questions:

\begin{enumerate}[(1)]
	\item \textit{Is the theoretical method I have developed valid?}, or in the case of reusing an existing method: \textit{Have I selected and configured appropriately the theoretical method and therefore, the \code implementing this method?}
	\item \textit{Is the \code implementing this method correct?}
\end{enumerate}

\noindent
While the first question concerns the \textit{validation} of software (``Am I building the right product''), the second one addresses its \textit{verification} (``Am I building the product right'', \cf~\cite{Boehm1984}).
Thus, validation is about the adequate (proper use), and verification about the correct (absence of bugs), functional behavior of software such as the \codes in CMS.
Hence, both validation and verification of CMS \codes are required to obtain trustworthy results from calculations.
Otherwise, the \codes pose a potential major threat to the validity of the experiments and research results, as any inadequate or incorrect \code refutes these results.
However, validating and verifying \codes in computational materials science poses challenges, which we will discuss in the next section.

\subsection{Challenges for Validating and Verifying CMS \Codes}
\label{subsec:challenges}

\noindent
In the following, we discuss major challenges for the verification and validation of scientific software, which - based on our experience in (research) software engineering - are caused by the described factors: experimental nature and complexity of \codes, and complexity of the data processed by them.

\subsubsection{Lack of Precise Oracles}
As scientists use computational calculations to explore new ideas and theoretical methods, the outcome of a calculation is generally not known at all, or at least not precisely known a priori~\cite{CarverKSP07,KellySM2011,Hannay2009}.
Other reasons for this are the complexity of the calculations, and the fact that the calculations may return a range of different answers, which makes it difficult for scientists to predict the outcome~\cite{Kanewala2014}.

This causes uncertainty about calculation results, as there is no precise notion of their correctness. Consequently, the same is true for the software used to explore such new ideas and methods, which prohibits precise oracles to be defined for quality assurance techniques such as software testing.

Consider, for example, a CMS scientist who performs one of the following calculations. She simulates existing materials with well-known properties on a new implementation (\code) of an existing, altered, or completely new method, or she simulates unknown materials on an existing \code of established methods, or even does so in bulk to explore the huge space of unknown potential materials.
To be more specific, considering the example presented previously (see Listing~\ref{listing:input}), estimating the impact of varying input parameters such as \codeGreen{cut} on the results may be difficult.
In all cases, the expected result of the calculation is not known and cannot be predicted a priori.

In contrast, the result obtained for a specific property of an input material is \textit{likely} to be correct if the calculated property value is statistically similar to that of other well-known materials of the same class, assuming we can  classify the material.
Outliers, in turn, may indicate one of the following cases:
\begin{enumerate}[(i)]
	\item Discovery of a highly interesting material.
	In this positive case, a material has been discovered, whose properties are different from existing materials of the same class.
	\item Faulty theoretical method and/or parameters.
	In this negative case, a scientist either made a mistake when developing a new method, which manifests in its implementation (\code), or used a nonsensical combination of method and parameters when using an existing method and \code. Here, the \code is, or may be, free of bugs.
	\item A bug in the software (\code). This is the other negative case, in which the \code contains a bug that caused the faulty results. More specifically, the \code is a faulty implementation of a valid theoretical method. 	
\end{enumerate}

This perspective gives rise to developing \textit{statistical oracles}, which judge the plausibility of computational results and provide corresponding feedback to scientists, which in turn establishes confidence in these results.
Non-plausible results, which are expected to be rare, need to be inspected manually and classified according to the cases (i), (ii), and~(iii).
The use of such statistical oracles is conceivable in quality assurance techniques such as software testing. However, from a software engineering point of view, testing mainly focuses on precise oracles and assertions, so that state-of-the-art and state-of-the-practice testing approaches, or even test-driven development, cannot be directly applied here. Consequently, quality assurance techniques such as systematic testing known from software engineering~\cite{Hannay2009,Wilson2014} are rarely adopted for scientific software~\cite{Kanewala2014}.
Therefore, increasing the confidence of scientists in computational results requires quality assurance techniques which can be applied to scientific software packages \textit{\mbox{a-posteriori}} and in an automated manner~\cite{Wilson2014}.

This results in the following challenges for leveraging statistical oracles in testing of scientific software:
\begin{itemize}
	\item Which methods and techniques shall be used to provide a statistical oracle?
	\item How can such methods reliably judge the success and potential failure of a set of executed tests?
\end{itemize}

\subsubsection{Large Configuration Space}
As discussed above, the experimental nature of scientific software (\codes) typically results in a large configuration space. This comprises the selection of algorithms provided by a software, as well as fine-tuning the selected algorithm through parameters, which in turn results in high variability and a large number of options for executing the software~\cite{RemmelPBE12, VilkomirSPC08}. At the same time, the choice of configuration influences the calculation results~\cite{Cashman:2018}.

An example from CMS is the calculation of single-particle band gaps, for which \exciting can be customized to perform a calculation that is further configured by a set of parameters (see Section~\ref{subsec:problem}). In the context of \Nomad, 40+~\codes such as \exciting are used, which multiplies the variability that CMS scientists have to cope with.

This variability challenges scientists to select and configure appropriate \codes for calculations.
As the selection and configuration of \codes can greatly influence the calculation results, CMS scientists should be supported \textit{\mbox{a-priori}} and in an automated manner during this process.
Such support should guide scientists in implementing a method to prevent the introduction of basic faults, before a calculation is conducted. It therefore promotes the validation of the configured methods/\codes and of the conducted calculations. In CMS, such support may suggest to scientists the use of trusted \codes and methods (including parameters) for specific materials and/or properties that are of interest for a specific calculation.
For example, a recommendation may be to use an all-electron \code and a self-interaction corrected exchange-correlation function to properly account for electron-electron interactions for a heavy material like cerium.

Moreover, the large configuration space and the corresponding variability of \codes also challenges the validation and verification of these \codes through testing, in that it is infeasible to test \textit{all} possible configurations. The large configuration space impedes manual identification of test cases and thereby of configurations to be tested (\cf~\cite{Kanewala2014, VilkomirSPC08, RemmelPBE12}). Thus, an automated sampling of the configuration space to identify representative configurations to be tested is required. In general, this constitutes a combinatorial interaction testing problem~\cite{Yilmaz2014} while a solution for this problem has to be tailored to the CMS domain.
Consequently, coping with the large configuration space requires automated support for scientists in using \codes, as well as intelligent testing techniques, which account for the following challenges:

\begin{itemize}
	\item What are appropriate sampling strategies for selecting a subset of scientific computations (i.e., a combination of \code, \code configuration, and input data in CMS) that are likely to reveal a failure in a scientific software?
	\item How to exploit results of previous calculations and test runs of \codes, to automatically determine the required support for scientists?
	Particularly, how to exploit statistical information from automated testing, to suggest methods and corresponding \codes (including configuration parameters) to scientists for a specific calculation?
\end{itemize}

\subsubsection{Large-Scale, Heterogeneous Data}
Scientific software often processes large-scale, heterogeneous data, e.g., in climate research~\cite{Easterbrook2010}, and in CMS~\cite{DraxlS2018} where software operates on data up to several TB in size and encoded in different \code-specific formats that are mostly neither formalized nor well-documented. 
Thus, calculations using results of multiple \codes in \Nomad require pre- and post-processing steps to transform input/output data between the normalized Archive format (\cf~Section~\ref{sec:cms}) and the \code-specific formats, to integrate machine learning, or for visualization.
For instance, parsers (one for each \code-specific format) and normalizers are used to translate \code-specific input/output data of a \code for storage in the Archive and future use.
Consequently, \codes implementing theoretical methods are embedded in a \textit{workflow}, together with programs implementing such pre- and post-processing steps.
One workflow example is a machine learning approach applied to properties computed by multiple \codes over many materials, to find predictors for a specific materials property. Here, the properties have to be computed, parsed, and normalized for the learning.

Consequently, validation and verification have to address the whole workflow. Otherwise, a faulty pre- or post-processing step might introduce faults into the data causing either wrong calculations by the (bug-free) \codes, or wrong presentations and interpretations of the results by scientists. Hence, a fault might be located in any part of a data-analysis workflow (\cf~\Nomad workflow in Section~\ref{sec:cms}).

Thus, the selection of test data (including the pre- and post-processing steps) for testing workflows is crucial.
For instance, considering a workflow that classifies materials based on their electrical resistivity and conductivity, tests should cover calculation data from different \codes implying different methods, unit systems, respective parser and normalizer chains, as well as representatives from different classes of materials (e.g., super-, semi-, non-, conductors).

This heterogeneity, together with the scale of the data, results in high variability at the data level (in addition to the variability of the \codes discussed in the context of large configuration spaces), which challenges the validation and verification of data-analysis workflows:

\begin{itemize}
	\item How to identify and sample valid/realistic test data for \codes and workflows that likely reveal a failure?
	\item How to improve the quality of the pre-/post-processing steps that handle large-scale, heterogeneous data?
\end{itemize}

\subsubsection{Global Software Development}
In CMS, scientists across the globe explore theoretical methods and develop \codes. An ecosystem of several hundred scientists and research groups has emerged around \Nomad, fostering
reuse of data for new and reproducing calculations,
reuse of \codes in workflows,
and
development of new \codes based on existing ones.

However, reuse is often kept implicit, e.g., for lack of common workflow descriptions~\cite{amstutz_common_2016}. For instance, relations between calculations do exist, but such relations often have to be deduced from common practices such as a commonly used layout of directories. For example, to derive elasticity properties for a material with \exciting, a series of simulations with varying forces acting on the simulated material has to be performed~\cite{Golesorkhtabar2013}. Only an analysis of all these simulations allows scientists to derive the desired elastic constants.
However, the intent behind the series of simulations is not always formalized. From the perspective of \Nomad, or data reuse in general, the parameter study's
relations between those simulations and their underlying intent have to be deduced. Even originally unrelated simulations from different \codes could be used in a parameter study, provided one identifies respective data based on comparable methods and parameters.

Moreover, \codes are sometimes not well-documented -- with regards to any and all levels of documentation, e.g., requirements, system modeling, architectural design, maintenance guidelines, and user documentation -- or no longer maintained, their data format may not be formalized, and the corresponding parser may only produce partial parses of the format. Finally, the quality of a \code might be unknown or the quality might differ, depending on the degree to which quality assurance techniques such as testing are adopted.
These aspects are caused by general issues of global software development concerning knowledge, project, and process management~\cite{HerbslebM2001}, and they challenge the validation and verification of \codes that are (re)used by scientists other than the scientists developing the \codes:

\begin{itemize}
	\item How to validate and verify third-party \code that is not well-documented, not sufficiently tested, and whose data format is not formalized? How to achieve trustworthy workflows that use data from different sources in different~\codes?
	\item How to extract and mine relations between calculations to leverage integration testing and to generally improve quality/trust levels of \codes, for instance, by externalizing assurances obtained for reused \codes?
\end{itemize}

\section{Directions for Future Research}
\label{sec:research}

\subsubsection{Lack of Precise Oracles}\label{subsubsec:lack-of-oracles}
Currently, the confidence about scientific research results in the CMS domain is addressed by scientific workflow systems using the notion of \textit{provenance}, since all executions of \codes and the corresponding input and output data are documented in \Nomad\footnote{\url{https://metainfo.nomad-coe.eu/nomadmetainfo_public/archive.html}}.
A recent community effort~\cite{Lejaeghere2016} has, for the first time, assessed and compared the quality of DFT results computed by several \codes for a set of materials. More recently, the effect of computational parameters have been systematically assessed, involving four different \codes~\cite{Carbogno2018}. The goal here is to automate the collection of workflow metadata to enable reproducibility of scientific results.

Based on the available \Nomad data, the next step will be to define a notion of a \textit{statistical oracle}, which uses statistical methods for identifying the correctness of a computational calculation~\cite{KanewalaB13}.
Unlike usual oracles used in software testing such as oracles derived from requirement specifications or models, gold standard oracles, or human oracles, the decision of a statistical oracle as envisioned is, by definition, not always correct. To apply statistical methods, results in the neighborhood of the computational calculation need to be investigated. Chan et al.~\cite{ChanCHT09} provide a general algorithm based on mesh specifications and machine learning for this problem. However, defining the neighborhood in CMS requires looking at the used materials, \codes including its parameters, and computational environment. Furthermore, the selection of appropriate heuristics, which keep the oracle's failure at a minimum, is an open problem that has been little researched in general~\cite{KanewalaB13} and needs to be tailored to data-driven CMS. Finally, if the neighborhood of a calculation is not available in \Nomad, specific computational calculations can be provided by mutation sensitivity testing~\cite{HookK09MutationSensitivityTesting} and modeling as well as approximation techniques of the input space~\cite{VilkomirSPC08}.

Beyond this, it will be very interesting to apply the concept of  metamorphic  testing~\cite{Chen1998metamorphic} that is specifically designed to test software without an oracle~\cite{BarrHMSY15,SeguraFSC16}. The idea is to identify and refine a set of metamorphic relations between the software inputs and outputs. Just to give an abstract example, for a square root function \texttt{sqrt(x)} the relation \texttt{x=sqrt(x)*sqrt(x)} should hold under reasonable floating point accuracy assumptions. Identifying such relations is highly domain dependent. However, automatic techniques based on machine learning have been proposed~\cite{GotliebB03,KanewalaB13,MurphyKHW08,XieHMKXC11} and successfully applied to the bio-medical~\cite{Chen2009innovative} and particle physics~\cite{DingZ16} domains. Transferring the concept of metamorphic testing requires domain expertise since the identified relations need to be understood and explained. The explainability of the relations is a primary challenge. However, it is also a significant opportunity for the CMS community since the scientists might learn hidden relations from their \codes which were previously unknown. This may strengthen the understanding and help to refine the underlying theories.

\subsubsection{Large Configuration Space}

\Codes in CMS are used by selecting a desired method and by fine-tuning this method through a set of parameters. Thus, there is not \textit{the} perfect implementation, but each \code is actually a tool box that can be instantiated in a huge number of variants. In sum, this leads to a combinatorial explosion of possible computations, and testing all of them is infeasible. Instead, appropriate sampling strategies are required which are effective and efficient at the same time. Possible scenarios for first tests are, for instance, to stay within the same \code family and vary, for a given method, the parameter space; or select the best possible (fully converged) calculations from different \codes.

The steps for future research to deal with the large configuration space when verifying and validating CMS  \codes and calculations require effective methods for configuration space sampling and automated test input generation.
Concerning the sampling of suitable input data for generated test cases, one idea~\cite{RemmelPBE12} is to apply combinatorial testing and test case selection techniques, which have been exploited in software product line (SPL) engineering~\cite{SarkarGSAC15,ThumAKSS14}. The goal of these techniques is to select a promising subset of product variants when testing all variants of an SPL is infeasible. However, SPL engineering focuses on testing interactions of features, which may be present in a product variant, or not. In contrast, the configuration space of CMS software comprises a set of non-boolean parameters, which demand different coverage metrics and sampling strategies.
Furthermore, our hypothesis is that the size of the configuration space exceeds the configuration space of very large existing SPLs such as the Linux kernel. As a result, we have to enrich the sampling strategies with modeling techniques for the input space as proposed by Vilkomir et al.~\cite{VilkomirSPC08}.
Another direction for future research is to use a recommender system exploiting statistical information obtained from existing computational calculations in the \Nomad Repository. Passing and non-passing test cases will be classified statistically to provide valuable information about the adequacy of \code configurations. Our aim is to exploit this information to derive a recommender system that assists scientists in configuring \codes for their specific needs.

\subsubsection{Large-Scale, Heterogeneous Data}
There are two aspects of the data diversity problem in CMS.
First, we have different representations of the same information, for instance, different file formats, layouts of matrices, units, etc.
Second, \codes provide different kinds of information, for instance, \codes specializing in electronic properties vs. \codes specializing in thermal properties.
The former problem can be solved by finding the right abstractions, the latter by defining relations between properties (e.g., identifying generalizations, categories of properties, or associations). Both aspects can be tackled by formal data models.

Modeling data has a long history in computer science, and has different methods in different \emph{technical spaces}~\cite{Ivanov2002} such as schemas for data exchange (XML, JSON), relational algebra in databases, ontologies in semantic web, or formal grammars and meta-models in computer languages. Applications often require transforming data from a representation in one space to a representation in another (e.g., reading data from a database organized in tables and sending it over the internet in hierarchically nested JSON format). The scale of the data increases the problem since specialized technologies have to be combined. For instance, search engines, distributed computing platforms, and nosql-databases have to work hand in hand. Each technology potentially requires its own specialized data representation. To cope with this, data must be modeled at a level that is independent from concrete technical spaces.

The CMS domain (or scientific software in general) presents a further challenge, since most existing methods for formally defining data types fall short as they neglect the nature of scientific data and offer no or insufficient support for vectors, matrices, tensors, their dimensions, and units. Therefore, \mbox{\Nomad} defines its own schema language \emph{meta-info}~\cite{Ghiringhelli2016} that is independent of the concrete data representation (e.g., text files, HDF5 files, or databases). In all its representations, data retains its inherent structure and types as defined in meta-info. Furthermore, meta-info categorizes properties into sections, and defines relations between properties and their categories. Some of the meta-info is common and shared by many \codes, some definitions are \code specific.

This formal model of CMS data can foster the quality of \codes and workflows in several ways, e.g., by applying methods from model-based testing. %
First, a formal model can support generating realistic large-scale test data and asserting test coverage with respect to the input of \codes.
Secondly, it is a formal definition of the possible data space. Constraints defined at the meta-info level can be used to automatically assert the plausibility of calculated properties.
Finally, it can automatize the development of mappings between technical spaces (i.e., parsers and normalizers) by declaratively defining mappings, from which operational transformations are automatically derived. This avoids error-prone manual implementations of parsers and normalizers.

\subsubsection{Global Software Development}
Global software development challenges to the verification and validation of \codes (re)used in CMS studies pertain mainly to two factors:
\begin{enumerate*}[(i)]
	\item The large and diverse development ecosystem, which produces \codes that differ in quality (e.g., levels of documentation and testing);
	\item the lack of explication of intent when combining multiple calculations and \codes in workflows.
\end{enumerate*}

Efforts to consolidate the diversity of the ecosystem in terms of software quality will have to be implemented as community processes. \Code development should adopt best practices of software engineering~\cite{Wilson2014,hastings_ten_2014}. These practices must be adapted to the needs of CMS, for instance, in regard to testing (\cf~Section~\ref{subsubsec:lack-of-oracles}). Similar efforts have been made in astronomy\footnote{\url{https://eas.unige.ch//EWASS2017/session.jsp?id=SS16}}. Such efforts will ease the integration and testing of \codes developed by other scientists in workflows.

Despite the use of workflow systems in CMS %
(\cf~Section~\ref{subsubsec:lack-of-oracles}), metadata explicating the intent behind a parametrization and combination of calculations/\codes within a single study is often missing. Therefore, any intent can only be deduced from potentially interrelated, non-formalized information such as directory structures.
To explicate intent, future efforts should develop and apply requirements for formalized metadata, for instance, by using the Common Workflow Language~\cite{amstutz_common_2016}, a specification for portable and scalable workflow descriptions with dedicated metadata. Similarly to efforts regarding the development ecosystem as such, this must be achieved through a standardization process within the CMS community. Additionally, automatic methods for the discovery of intentional process models based on Hidden Markov Models~\cite{khodabandelou_supervised_2013,khodabandelou_unsupervised_2014} can be adapted to mine implicit relations between calculations/\codes. These models can guide the integration testing of data analysis workflows.

\section{Conclusions}
In this paper we discussed challenges for the validation and verification of scientific software in computational materials science (CMS). We conclude that most of the problems are similar to other domains \cite{KanewalaB13,Nanthaamornphong2017,Dubois12,Clune2014testing,HookK09,KellyTH11,Kelly15,MorrisS09,ReupkeSRC88} and solution principles derived for the CMS domain might be generalizable to other domains. However, the effort of the CMS community to provide results of their computational experiments in the \Nomad Repository~\cite{DraxlS2018} based on the FAIR principle \cite{wilkinson_fair_2016} provides a significant opportunity for fundamental research on validation and verification of scientific software. For instance, based on the \Nomad data, novel strategies to tackle the oracle problem, can be developed. With this research, we envision trust levels for \codes so that scientists increase their trust in \codes to obtain trustworthy, reproducible calculations and research results.


\begin{thebibliography}{10}
	\providecommand{\url}[1]{#1}
	\csname url@samestyle\endcsname
	\providecommand{\newblock}{\relax}
	\providecommand{\bibinfo}[2]{#2}
	\providecommand{\BIBentrySTDinterwordspacing}{\spaceskip=0pt\relax}
	\providecommand{\BIBentryALTinterwordstretchfactor}{4}
	\providecommand{\BIBentryALTinterwordspacing}{\spaceskip=\fontdimen2\font plus
		\BIBentryALTinterwordstretchfactor\fontdimen3\font minus
		\fontdimen4\font\relax}
	\providecommand{\BIBforeignlanguage}[2]{{%
			\expandafter\ifx\csname l@#1\endcsname\relax
			\typeout{** WARNING: IEEEtran.bst: No hyphenation pattern has been}%
			\typeout{** loaded for the language `#1'. Using the pattern for}%
			\typeout{** the default language instead.}%
			\else
			\language=\csname l@#1\endcsname
			\fi
			#2}}
	\providecommand{\BIBdecl}{\relax}
	\BIBdecl
	
	\bibitem{Carver2016se4sience}
	J.~C. Carver, N.~P.~C. Hong, and G.~K. Thiruvathukal, \emph{Software
		Engineering for Science}.\hskip 1em plus 0.5em minus 0.4em\relax CRC Press,
	2016.
	
	\bibitem{SandersK08}
	R.~Sanders and D.~Kelly, ``Dealing with risk in scientific software
	development,'' \emph{{IEEE} Software}, vol.~25, no.~4, pp. 21--28, 2008.
	
	\bibitem{Miller1856}
	G.~Miller, ``A scientist{\textquoteright}s nightmare: Software problem leads to
	five retractions,'' \emph{Science}, vol. 314, no. 5807, pp. 1856--1857, 2006.
	
	\bibitem{Hannay2009}
	J.~E. Hannay, C.~MacLeod, J.~Singer, H.~P. Langtangen, D.~Pfahl, and G.~Wilson,
	``How do scientists develop and use scientific software?'' in \emph{ICSE
		Workshop on Software Engineering for Computational Science and Engineering},
	ser. SECSE.\hskip 1em plus 0.5em minus 0.4em\relax IEEE, 2009, pp. 1--8.
	
	\bibitem{Nguyen-HoanFS10}
	L.~Nguyen{-}Hoan, S.~Flint, and R.~Sankaranarayana, ``A survey of scientific
	software development,'' in \emph{International Symposium on Empirical
		Software Engineering and Measurement}.\hskip 1em plus 0.5em minus 0.4em\relax
	ACM, 2010, pp. 12:1--12:10.
	
	\bibitem{HeatonC15}
	D.~Heaton and J.~C. Carver, ``Claims about the use of software engineering
	practices in science: {A} systematic literature review,'' \emph{Information
		{\&} Software Technology}, vol.~67, pp. 207--219, 2015.
	
	\bibitem{Storer17}
	T.~Storer, ``Bridging the chasm: {A} survey of software engineering practice in
	scientific programming,'' \emph{{ACM} Comput. Surv.}, vol.~50, no.~4, pp.
	47:1--47:32, 2017.
	
	\bibitem{JohansonH18}
	A.~N. Johanson and W.~Hasselbring, ``Software engineering for computational
	science: Past, present, future,'' \emph{Computing in Science and
		Engineering}, vol.~20, no.~2, pp. 90--109, 2018.
	
	\bibitem{KellyHS09}
	D.~Kelly, D.~Hook, and R.~Sanders, ``Five recommended practices for
	computational scientists who write software,'' \emph{Computing in Science and
		Engineering}, vol.~11, no.~5, pp. 48--53, 2009.
	
	\bibitem{Wilson2014}
	G.~Wilson, D.~A. Aruliah, C.~T. Brown, N.~P. Chue~Hong, M.~Davis, R.~T. Guy,
	S.~H.~D. Haddock, K.~D. Huff, I.~M. Mitchell, M.~D. Plumbley, B.~Waugh, E.~P.
	White, and P.~Wilson, ``Best practices for scientific computing,'' \emph{PLOS
		Biology}, vol.~12, no.~1, pp. 1--7, 01 2014.
	
	\bibitem{Segal05}
	J.~Segal, ``When software engineers met research scientists: {A} case study,''
	\emph{Empirical Software Eng.}, vol.~10, no.~4, pp. 517--536, 2005.
	
	\bibitem{Segal08}
	------, ``Scientists and software engineers: {A} tale of two cultures,'' in
	\emph{Workshop of the Psychology of Programming Interest Group}, 2008, p.~6.
	
	\bibitem{HattonR94}
	L.~Hatton and A.~Roberts, ``How accurate is scientific software?'' \emph{{IEEE}
		Trans. Software Eng.}, vol.~20, no.~10, pp. 785--797, 1994.
	
	\bibitem{Clune2014testing}
	T.~Clune, M.~Rilee, and D.~Rouson, ``Testing as an essential process for
	developing and maintaining scientific software,'' in \emph{The 2nd Workshop
		on Sustainable Software for Science: Practices and Experiences}, 2014.
	
	\bibitem{KellyGS11}
	D.~Kelly, R.~Gray, and Y.~Shao, ``Examining random and designed tests to detect
	code mistakes in scientific software,'' \emph{J. Comput. Science}, vol.~2,
	no.~1, pp. 47--56, 2011.
	
	\bibitem{KellyTH11}
	D.~Kelly, S.~Thorsteinson, and D.~Hook, ``Scientific software testing: Analysis
	with four dimensions,'' \emph{Softw.}, vol.~28, no.~3, pp. 84--90, 2011.
	
	\bibitem{LaneG12}
	P.~C.~R. Lane and F.~Gobet, ``A theory-driven testing methodology for
	developing scientific software,'' \emph{J. Exp. Theor. Artif. Intell.},
	vol.~24, no.~4, pp. 421--456, 2012.
	
	\bibitem{Kelly15}
	D.~Kelly, ``Scientific software development viewed as knowledge acquisition:
	Towards understanding the development of risk-averse scientific software,''
	\emph{Journal of Systems and Software}, vol. 109, pp. 50--61, 2015.
	
	\bibitem{Kanewala2014}
	U.~Kanewala and J.~M. Bieman, ``Testing scientific software: A systematic
	literature review,'' \emph{Information and Software Technology}, vol.~56,
	no.~10, pp. 1219--1232, 2014.
	
	\bibitem{SmithKRY04}
	M.~C. Smith, R.~L. Kelsey, J.~M. Riese, and G.~A. Young, ``Creating a flexible
	environment for testing scientific software,'' in \emph{Enabling Technologies
		for Simulation Science\,VIII,\,SPIE\,5423}, 2004, pp. 288--296.
	
	\bibitem{Dubois12}
	P.~F. Dubois, ``Testing scientific programs,'' \emph{Computing in Science and
		Engineering}, vol.~14, no.~4, pp. 69--73, 2012.
	
	\bibitem{CoxH1999}
	M.~Cox and P.~Harris, ``Design and use of reference data sets for testing
	scientific software,'' \emph{Analytica Chimica Acta}, vol. 380, no.~2, pp.
	339--351, 1999.
	
	\bibitem{Nanthaamornphong2017}
	A.~Nanthaamornphong and J.~C. Carver, ``Test-driven development in scientific
	software: a survey,'' \emph{Software Quality Journal}, vol.~25, no.~2, pp.
	343--372, Jun 2017.
	
	\bibitem{CluneR11}
	T.~L. Clune and R.~Rood, ``Software testing and verification in model
	development,'' \emph{{IEEE} Software}, vol.~28, no.~6, pp. 49--55, 2011.
	
	\bibitem{CarverKSP07}
	J.~C. Carver, R.~P. Kendall, S.~E. Squires, and D.~E. Post, ``Software
	development environments for scientific and engineering software: {A} series
	of case studies,'' in \emph{29th International Conference on Software
		Engineering {(ICSE)}}.\hskip 1em plus 0.5em minus 0.4em\relax {IEEE}, 2007,
	pp. 550--559.
	
	\bibitem{HookK09}
	D.~Hook and D.~Kelly, ``Testing for trustworthiness in scientific software,''
	in \emph{{ICSE} Workshop on Software Engineering for Computational Science
		and Engineering, {SE-CSE}}.\hskip 1em plus 0.5em minus 0.4em\relax {IEEE},
	2009, pp. 59--64.
	
	\bibitem{DraxlS2018}
	C.~Draxl and M.~Scheffler, ``Nomad: The fair concept for big data-driven
	materials science,'' \emph{MRS Bulletin}, vol.~43, no.~9, pp. 676--682, 2018.
	
	\bibitem{exciting}
	A.~Gulans, S.~Kontur, C.~Meisenbichler, D.~Nabok, P.~Pavone, S.~Rigamonti,
	S.~Sagmeister, U.~Werner, and C.~Draxl, ``exciting: a full-potential
	all-electron package implementing density-functional theory and many-body
	perturbation theory,'' \emph{Journal of Physics: Condensed Matter}, vol.~26,
	no.~36, p. 363202, 2014.
	
	\bibitem{abinit}
	X.~Gonze, F.~Jollet, F.~Abreu~Araujo, D.~Adams, B.~Amadon, T.~Applencourt,
	C.~Audouze, J.~M. Beuken, J.~Bieder, A.~Bokhanchuk, E.~Bousquet, F.~Bruneval,
	D.~Caliste, M.~C\^ot\'e, F.~Dahm, F.~Da~Pieve, M.~Delaveau, M.~Di~Gennaro,
	B.~Dorado, C.~Espejo, G.~Geneste, L.~Genovese, A.~Gerossier, M.~Giantomassi,
	Y.~Gillet, D.~R. Hamann, L.~He, G.~Jomard, J.~Laflamme~Janssen, S.~Le~Roux,
	A.~Levitt, A.~Lherbier, F.~Liu, I.~Luka{\v c}evi\'c, A.~Martin, C.~Martins,
	M.~J.~T. Oliveira, S.~Ponc\'e, Y.~Pouillon, T.~Rangel, G.~M. Rignanese, A.~H.
	Romero, B.~Rousseau, O.~Rubel, A.~A. Shukri, M.~Stankovski, M.~Torrent, M.~J.
	Van~Setten, B.~Van~Troeye, M.~J. Verstraete, D.~Waroquiers, J.~Wiktor, B.~Xu,
	A.~Zhou, and J.~W. Zwanziger, ``Recent developments in the {{ABINIT}}
	software package,'' \emph{Computer Physics Communications}, vol. 205, pp.
	106--131, Aug. 2016.
	
	\bibitem{wilkinson_fair_2016}
	M.~D. Wilkinson, M.~Dumontier, I.~J. Aalbersberg, G.~Appleton, M.~Axton,
	A.~Baak, N.~Blomberg, J.-W. Boiten, L.~B. {da Silva Santos}, P.~E. Bourne,
	J.~Bouwman, A.~J. Brookes, T.~Clark, M.~Crosas, I.~Dillo, O.~Dumon,
	S.~Edmunds, C.~T. Evelo, R.~Finkers, A.~{Gonzalez-Beltran}, A.~J.~G. Gray,
	P.~Groth, C.~Goble, J.~S. Grethe, J.~Heringa, P.~A.~C. {'t Hoen}, R.~Hooft,
	T.~Kuhn, R.~Kok, J.~Kok, S.~J. Lusher, M.~E. Martone, A.~Mons, A.~L. Packer,
	B.~Persson, P.~{Rocca-Serra}, M.~Roos, R.~{van Schaik}, S.-A. Sansone,
	E.~Schultes, T.~Sengstag, T.~Slater, G.~Strawn, M.~A. Swertz, M.~Thompson,
	J.~{van der Lei}, E.~{van Mulligen}, J.~Velterop, A.~Waagmeester,
	P.~Wittenburg, K.~Wolstencroft, J.~Zhao, and B.~Mons,
	``\BIBforeignlanguage{en}{The {{FAIR Guiding Principles}} for scientific data
		management and stewardship},'' \emph{\BIBforeignlanguage{en}{Scientific
			Data}}, vol.~3, p. 160018, Mar. 2016.
	
	\bibitem{RodriguesPela+2018}
	R.~Rodrigues~Pela, A.~Gulans, and C.~Draxl, ``The lda-1/2 method applied to
	atoms and molecules,'' \emph{Journal of Chemical Theory and Computation},
	vol.~14, no.~9, pp. 4678--4686, 2018, pMID: 30119607.
	
	\bibitem{Cashman:2018}
	M.~Cashman, M.~B. Cohen, P.~Ranjan, and R.~W. Cottingham, ``Navigating the
	maze: The impact of configurability in bioinformatics software,'' in
	\emph{33rd International Conference on Automated Software Engineering}, ser.
	ASE.\hskip 1em plus 0.5em minus 0.4em\relax ACM, 2018, pp. 757--767.
	
	\bibitem{Boehm1984}
	B.~W. Boehm, ``Verifying and validating software requirements and design
	specifications,'' \emph{IEEE Software}, vol.~1, no.~1, pp. 75--88, 1984.
	
	\bibitem{KellySM2011}
	D.~Kelly, S.~Smith, and N.~Meng, ``Software engineering for scientists,''
	\emph{Computing in Science Engineering}, vol.~13, no.~5, pp. 7--11, 2011.
	
	\bibitem{RemmelPBE12}
	H.~Remmel, B.~Paech, P.~Bastian, and C.~Engwer, ``System testing a scientific
	framework using a regression-test environment,'' \emph{Computing in Science
		and Engineering}, vol.~14, no.~2, pp. 38--45, 2012.
	
	\bibitem{VilkomirSPC08}
	S.~A. Vilkomir, W.~T. Swain, J.~H. Poore, and K.~T. Clarno, ``Modeling input
	space for testing scientific computational software: {A} case study,'' in
	\emph{8th International Conference on Computational Science (ICCS)}, ser.
	LNCS, vol. 5103.\hskip 1em plus 0.5em minus 0.4em\relax Springer, 2008, pp.
	291--300.
	
	\bibitem{Yilmaz2014}
	C.~Yilmaz, S.~Fouché, M.~B. Cohen, A.~Porter, G.~Demiroz, and U.~Koc, ``Moving
	forward with combinatorial interaction testing,'' \emph{Computer}, vol.~47,
	no.~2, pp. 37--45, 2014.
	
	\bibitem{Easterbrook2010}
	S.~M. Easterbrook, ``Climate change: A grand software challenge,'' in
	\emph{Proceedings of the FSE/SDP Workshop on Future of Software Engineering
		Research}, ser. FoSER '10.\hskip 1em plus 0.5em minus 0.4em\relax ACM, 2010,
	pp. 99--104.
	
	\bibitem{amstutz_common_2016}
	\BIBentryALTinterwordspacing
	B.~Chapman, J.~Chilton, M.~Heuer, A.~Kartashov, D.~Leehr, H.~M\'enager,
	M.~Nedeljkovich, M.~Scales, S.~{Soiland-Reyes}, and L.~Stojanovic, ``Common
	{{Workflow Language}}, v1.0,'' {Common Workflow Language working group},
	Specification, Jul. 2016. [Online]. Available:
	\url{https://w3id.org/cwl/v1.0/}
	\BIBentrySTDinterwordspacing
	
	\bibitem{Golesorkhtabar2013}
	R.~Golesorkhtabar, P.~Pavone, J.~Spitaler, P.~Puschnig, and C.~Draxl,
	``Elastic: A tool for calculating second-order elastic constants from first
	principles,'' \emph{Computer Physics Communications}, vol. 184, no.~8, pp.
	1861--1873, 2013.
	
	\bibitem{HerbslebM2001}
	J.~D. Herbsleb and D.~Moitra, ``Global software development,'' \emph{IEEE
		Software}, vol.~18, no.~2, pp. 16--20, March 2001.
	
	\bibitem{Lejaeghere2016}
	K.~Lejaeghere, G.~Bihlmayer, T.~Bj{\"o}rkman, P.~Blaha, S.~Bl{\"u}gel, V.~Blum,
	D.~Caliste, I.~E. Castelli, S.~J. Clark, A.~Dal~Corso, S.~de~Gironcoli,
	T.~Deutsch, J.~K. Dewhurst, I.~Di~Marco, C.~Draxl, M.~Du{\l}ak, O.~Eriksson,
	J.~A. Flores-Livas, K.~F. Garrity, L.~Genovese, P.~Giannozzi, M.~Giantomassi,
	S.~Goedecker, X.~Gonze, O.~Gr{\r a}n{\"a}s, E.~K.~U. Gross, A.~Gulans,
	F.~Gygi, D.~R. Hamann, P.~J. Hasnip, N.~A.~W. Holzwarth, D.~Iu{\c s}an, D.~B.
	Jochym, F.~Jollet, D.~Jones, G.~Kresse, K.~Koepernik, E.~K{\"u}{\c
		c}{\"u}kbenli, Y.~O. Kvashnin, I.~L.~M. Locht, S.~Lubeck, M.~Marsman,
	N.~Marzari, U.~Nitzsche, L.~Nordstr{\"o}m, T.~Ozaki, L.~Paulatto, C.~J.
	Pickard, W.~Poelmans, M.~I.~J. Probert, K.~Refson, M.~Richter, G.-M.
	Rignanese, S.~Saha, M.~Scheffler, M.~Schlipf, K.~Schwarz, S.~Sharma,
	F.~Tavazza, P.~Thunstr{\"o}m, A.~Tkatchenko, M.~Torrent, D.~Vanderbilt, M.~J.
	van Setten, V.~Van~Speybroeck, J.~M. Wills, J.~R. Yates, G.-X. Zhang, and
	S.~Cottenier, ``Reproducibility in density functional theory calculations of
	solids,'' \emph{Science}, vol. 351, no. 6280, 2016.
	
	\bibitem{Carbogno2018}
	C.~{Carbogno}, K.~{Thygesen}, B.~{Bieniek}, C.~{Draxl}, L.~{Ghiringhelli},
	A.~{Gulans}, O.~{Hofmann}, K.~{Jacobsen}, S.~{Lubeck}, J.~{Mortensen},
	M.~{Strange}, E.~{Wruss}, and M.~{Scheffler}, ``{Quality Control of Numerical
		Settings for DFT Calculations and Materials Databases},'' in \emph{APS
		Meeting Abstracts}, 2018, p. P12.003.
	
	\bibitem{KanewalaB13}
	U.~Kanewala and J.~M. Bieman, ``Techniques for testing scientific programs
	without an oracle,'' in \emph{Intl. Workshop on Software Engineering for
		Computational Science and Engineering}.\hskip 1em plus 0.5em minus
	0.4em\relax {IEEE}, 2013, pp. 48--57.
	
	\bibitem{ChanCHT09}
	W.~K. Chan, S.~Cheung, J.~C.~F. Ho, and T.~H. Tse, ``{PAT:} {A} pattern
	classification approach to automatic reference oracles for the testing of
	mesh simplification programs,'' \emph{Journal of Systems and Software},
	vol.~82, no.~3, pp. 422--434, 2009.
	
	\bibitem{HookK09MutationSensitivityTesting}
	D.~Hook and D.~Kelly, ``Mutation sensitivity testing,'' \emph{Computing in
		Science and Engineering}, vol.~11, no.~6, pp. 40--47, 2009.
	
	\bibitem{Chen1998metamorphic}
	T.~Y. Chen, S.~C. Cheung, and S.~M. Yiu, ``Metamorphic testing: a new approach
	for generating next test cases,'' HKUST-CS98-01, Department of Computer
	Science, Hong Kong~…, Tech. Rep., 1998.
	
	\bibitem{BarrHMSY15}
	E.~T. Barr, M.~Harman, P.~McMinn, M.~Shahbaz, and S.~Yoo, ``The oracle problem
	in software testing: {A} survey,'' \emph{{IEEE} Trans. Software Eng.},
	vol.~41, no.~5, pp. 507--525, 2015.
	
	\bibitem{SeguraFSC16}
	S.~Segura, G.~Fraser, A.~B. S{\'{a}}nchez, and A.~R. Cort{\'{e}}s, ``A survey
	on metamorphic testing,'' \emph{{IEEE} Trans. Software Eng.}, vol.~42, no.~9,
	pp. 805--824, 2016.
	
	\bibitem{GotliebB03}
	A.~Gotlieb and B.~Botella, ``Automated metamorphic testing,'' in \emph{27th
		International Computer Software and Applications Conference (COMPSAC)}.\hskip
	1em plus 0.5em minus 0.4em\relax {IEEE}, 2003, pp. 34--40.
	
	\bibitem{MurphyKHW08}
	C.~Murphy, G.~E. Kaiser, L.~Hu, and L.~Wu, ``Properties of machine learning
	applications for use in metamorphic testing,'' in \emph{Intl. Conf. on Softw.
		Engineering\,{\&}\,Knowledge Engineering}.\hskip 1em plus 0.5em minus
	0.4em\relax KSI, 2008, pp. 867--872.
	
	\bibitem{XieHMKXC11}
	X.~Xie, J.~W.~K. Ho, C.~Murphy, G.~E. Kaiser, B.~Xu, and T.~Y. Chen, ``Testing
	and validating machine learning classifiers by metamorphic testing,''
	\emph{Journal of Sys. and Softw.}, vol.~84, no.~4, pp. 544--558, 2011.
	
	\bibitem{Chen2009innovative}
	T.~Y. Chen, J.~W. Ho, H.~Liu, and X.~Xie, ``An innovative approach for testing
	bioinformatics programs using metamorphic testing,'' \emph{BMC
		bioinformatics}, vol.~10, no.~1, p.~24, 2009.
	
	\bibitem{DingZ16}
	J.~Ding and D.~Zhang, ``A machine learning approach for developing test oracles
	for testing scientific software,'' in \emph{Intl. Conference on Software
		Engineering and Knowledge Engineering}.\hskip 1em plus 0.5em minus
	0.4em\relax KSI, 2016, pp. 390--395.
	
	\bibitem{SarkarGSAC15}
	A.~Sarkar, J.~Guo, N.~Siegmund, S.~Apel, and K.~Czarnecki, ``Cost-efficient
	sampling for performance prediction of configurable systems {(T)},'' in
	\emph{30th International Conference on Automated Software Engineering,
		{ASE}}.\hskip 1em plus 0.5em minus 0.4em\relax {IEEE}, 2015, pp. 342--352.
	
	\bibitem{ThumAKSS14}
	T.~Th{\"{u}}m, S.~Apel, C.~K{\"{a}}stner, I.~Schaefer, and G.~Saake, ``A
	classification and survey of analysis strategies for software product
	lines,'' \emph{{ACM} Comput. Surv.}, vol.~47, no.~1, pp. 6:1--6:45, 2014.
	
	\bibitem{Ivanov2002}
	I.~Ivanov, J.~B{\'e}zivin, and M.~Aksit,
	``\BIBforeignlanguage{Undefined}{Technological spaces: An initial
		appraisal},'' 10 2002, pp. 1--6.
	
	\bibitem{Ghiringhelli2016}
	L.~M. Ghiringhelli, C.~Carbogno, S.~Levchenko, F.~Mohamed, G.~Huhs, M.~Lueders,
	M.~Oliveira, and M.~Scheffler, ``Towards a common format for computational
	material science data,'' 2016.
	
	\bibitem{hastings_ten_2014}
	J.~Hastings, K.~Haug, and C.~Steinbeck, ``Ten recommendations for software
	engineering in research,'' \emph{GigaScience}, vol.~3, Dec. 2014.
	
	\bibitem{khodabandelou_supervised_2013}
	G.~Khodabandelou, C.~Hug, R.~Deneck\`ere, and C.~Salinesi, ``Supervised
	intentional process models discovery using {{Hidden Markov}} models,'' in
	\emph{7th {{International Conference}} on {{Research Challenges}} in
		{{Information Science}} ({{RCIS}})}, 2013, pp. 1--11.
	
	\bibitem{khodabandelou_unsupervised_2014}
	------, ``Unsupervised {{Discovery}} of {{Intentional Process Models}} from
	{{Event Logs}},'' in \emph{Proceedings of the 11th {{Working Conference}} on
		{{Mining Software Repositories}}}, ser. {{MSR}} 2014.\hskip 1em plus 0.5em
	minus 0.4em\relax {ACM}, 2014, pp. 282--291.
	
	\bibitem{MorrisS09}
	C.~Morris and J.~Segal, ``Some challenges facing scientific software
	developers: The case of molecular biology,'' in \emph{Fifth International
		Conference on e-Science}.\hskip 1em plus 0.5em minus 0.4em\relax {IEEE},
	2009, pp. 216--222.
	
	\bibitem{ReupkeSRC88}
	W.~A. Reupke, E.~Srinivasan, P.~V. Rigterink, and D.~N. Card, ``The need for a
	rigorous development and testing methodology for medical software,'' in
	\emph{First Annual {IEEE} Symposium on Computer-Based Medical Systems
		(CBMS'88)}.\hskip 1em plus 0.5em minus 0.4em\relax {IEEE}, 1988, pp. 15--20.
	
\end{thebibliography}

\end{document}